\documentclass[a4paper,12pt]{article}

\usepackage[english]{babel}
\usepackage[utf8x]{inputenc}
\usepackage[T1]{fontenc}
\usepackage{authblk}
\usepackage{natbib}
\usepackage{here}
\usepackage{setspace}
\usepackage{multirow}
\usepackage{lmodern}
\usepackage{amsmath, amssymb}
\usepackage{bm}
\usepackage{pdflscape}
\usepackage{authblk}
\usepackage{color}
\usepackage[table]{xcolor}

\usepackage{soul}

\usepackage{booktabs}
\usepackage{afterpage}
\usepackage{textcomp}
\usepackage{float}
\usepackage{enumerate}
\usepackage{rotating}

\definecolor{babyblue}{rgb}{0.54, 0.81, 0.94}
\definecolor{babypink}{rgb}{0.96, 0.76, 0.76}

\usepackage[margin=1in]{geometry}

\usepackage{amsmath}
\usepackage{graphicx}
\usepackage[colorlinks=true, allcolors=blue]{hyperref}

\usepackage{caption}
\usepackage{subcaption}

\newcommand{\pg}{P{\'o}lya-Gamma }

\definecolor{Gray}{gray}{0.85}
\newcolumntype{a}{>{\columncolor{Gray}}c}
\newcolumntype{d}{>{\columncolor{white}}c}

\begin{document}

    \begin{center}
        \vspace*{1cm}
        \large
	    \texttt{spOccupancy}\textbf{: An R package for single-species, multi-species, and integrated spatial occupancy models}\\
         \normalsize
           \vspace{5mm}
	    Jeffrey W. Doser\textsuperscript{1, 2}, Andrew O. Finley\textsuperscript{1, 2}, Marc K{\'e}ry\textsuperscript{3}, Elise F. Zipkin\textsuperscript{2, 4}
         \vspace{5mm}
    \end{center}
    \small
         \textsuperscript{1}Department of Forestry, Michigan State University, East Lansing, MI, USA \\
         \textsuperscript{2}Ecology, Evolution, and Behavior Program, Michigan State University, East Lansing, MI, USA \\
         \textsuperscript{3}Swiss Ornithological Institute, Sempach, Switzerland \\
	     \textsuperscript{4} Department of Integrative Biology, Michigan State University, East Lansing, MI, USA \\
         \noindent \textbf{Corresponding Author}: Jeffrey W. Doser, email: doserjef@msu.edu; ORCID ID: 0000-0002-8950-9895 \\
         \noindent \textbf{Running Title}: \texttt{spOccupancy}: Spatial occupancy models in R
\section*{Abstract}

\begin{enumerate}
    \item Occupancy modeling is a common approach to assess species distribution patterns, while explicitly accounting for false absences in detection-nondetection data. Numerous extensions of the basic single-species occupancy model exist to model multiple species, spatial autocorrelation, and to integrate multiple data types. However, development of specialized and computationally efficient software to incorporate such extensions, especially for large data sets, is scarce or absent. 
    \item We introduce the \texttt{spOccupancy} \texttt{R} package designed to fit single-species and multi-species spatially-explicit occupancy models. We fit all models within a Bayesian framework using \pg data augmentation, which results in fast and efficient inference. \texttt{spOccupancy} provides functionality for data integration of multiple single-species detection-nondetection data sets via a joint likelihood framework. The package leverages Nearest Neighbor Gaussian Processes to account for spatial autocorrelation, which enables spatially-explicit occupancy modeling for potentially massive data sets (e.g., 1000s-100,000s of sites).
    \item \texttt{spOccupancy} provides user-friendly functions for data simulation, model fitting, model validation (by posterior predictive checks), model comparison (using information criteria and k-fold cross-validation), and out-of-sample prediction. We illustrate the package's functionality via a vignette, simulated data analysis, and two bird case studies. 
    \item The \texttt{spOccupancy} package provides a user-friendly platform to fit a variety of single and multi-species occupancy models, making it straightforward to address detection biases and spatial autocorrelation in species distribution models even for large data sets.  
\end{enumerate}

\noindent \textbf{Keywords}: Bayesian, data fusion, data integration, hierarchical model, imperfect detection, MCMC, occupancy model, spatial autocorrelation

\newpage 

\section{Introduction}

Understanding the processes that drive species distributions across space and time is a fundamental objective in ecology \citep{pulliam2000relationship}. Species distribution models (SDMs) are the primary tool used to study the spatial distributions of both individual species and entire communities \citep{guisan2000predictive}. Imperfect detection, or the failure to observe a species during sampling where it is in fact present, is a ubiquitous complication that must be addressed when modeling species distributions \citep{mackenzie2002, tyre2003improving}. Single-species occupancy models (SSOMs), a specialized type of SDM, explicitly incorporate the detection process separately from the latent species occurrence process using replicated detection-nondetection data \citep{mackenzie2002, tyre2003improving}. Multi-species occupancy models (MSOMs) are an extension to SSOMs that leverage detection-nondetection data from multiple species \citep{dorazio2005, gelfand2005modelling}. This approach views species-specific parameters as random effects arising from a common community-level distribution, which enables inferences at multiple scales (species, community) and leads to greater precision of species-specific effects and biodiversity metrics, all with fully propagated uncertainty \citep{dorazio2005}. 

While SSOMs and MSOMs were developed for use with only a single detection-nondetection data set, there is increasing interest in combining multiple data sources within a single statistical model to improve species distribution inferences (\citealt{isaac2020data}; Chapter 10; \citealt{keryRoyle2021}). Data integration (also referred to as data fusion) is a model-based approach that combines multiple data sources and accommodates different sampling processes among data sources \citep{miller2019recent}. Integrated occupancy models (IOMs) simultaneously analyze multiple detection-nondetection data sources to better estimate the latent process of interest and sources of uncertainty. These models are particularly attractive as there are numerous possible protocols to obtain detection-nondetection data sources across potentially vast spatial regions, such as autonomous recording units  \citep{doser2021integrating}. 

As detection-nondetection data sources increase in both spatial extent and number of observed locations, accounting for spatial autocorrelation becomes increasingly more important \citep{guelat2018}. Accommodating sources of spatial dependency among observations is key to delivering valid inferences about species distributions and has led to the development of spatial occupancy models \citep{johnson2013spatial}. Modeling spatial dependence via spatially structured random effects can improve predictive performance in occurrence probabilities across a region of interest \citep{wright2021spatial}. Including such random effects, however, is notoriously  computationally expensive and can easily lead to intolerable software run times when the number of locations used to define the spatial random effects becomes even moderately large (e.g., 100s-1000s of locations). This is particularly true when modeling detection-nondetection data as point locations in continuous space (i.e., in point-referenced spatial regression models) rather than as discrete units on a gridded study area (i.e., in areal spatial regression models; \citealt{banerjee2003}), as computational complexity increases in cubic order with the number of spatial locations. This so-called ``big N'' problem \citep{banerjee2012} renders even moderately large data sets computationally infeasible using common Bayesian software packages such as \texttt{Stan} \citep{carpenter2017}, \texttt{JAGS} \citep{plummer03}, and \texttt{NIMBLE} \citep{deValpine2017}. 

A paucity of user-friendly and computationally efficient software has so far limited adoption of spatial occupancy models in practice, as models may take weeks to run or simply not be possible to fit at all. In the context of SSOMs, \texttt{MARK} \citep{white1999program}, \texttt{PRESENCE} \citep{hines2006presence}, and the \texttt{R} package \texttt{unmarked} \citep{fiske2011} fit a variety of models for wildlife data using likelihood inference, but lack functionality to account for spatial autocorrelation. The \texttt{R} packages \texttt{stocc} \citep{johnson2013spatial}, \texttt{hSDM} \citep{hSDM}, \texttt{Rcppocc} \citep{clark2019}, and \texttt{ubms} \citep{kellner2021ubms} all fit spatial SSOMs using areal spatial models. However, these packages cannot fit MSOMs and may not be adequate for large data sets (e.g., thousands of locations). In a multi-species framework, the \texttt{R} package \texttt{HMSC} \citep{tikhonov2020joint} fits a wide range of spatially-explicit joint species distribution models allowing for estimation of complex correlations among species occurrence patterns, but this package fails to account for imperfect detection. Additionally, none of these packages facilitate model-based data integration, which together leads many practitioners to use Bayesian programming languages such as \texttt{JAGS}, \texttt{NIMBLE} and \texttt{Stan} to fit SSOMs, MSOMs, or IOMs. While these programming languages are incredibly flexible for defining specialized models, they are often not optimized for efficient computation involving dense covariance matrices and may pose a steep learning curve for novice users. 

Here we present \texttt{spOccupancy}, an \texttt{R} package that fits SSOMs, MSOMs, and IOMs that can accommodate spatial autocorrelation in potentially massive data sets. We fit spatial SSOMs, MSOMs, and IOMs in a point-referenced framework using either Gaussian processes or Nearest Neighbor Gaussian Processes (NNGPs; \citealt{datta2016hierarchical, finley2019efficient}). NNGPs use local information from a small set of nearest neighbors to closely approximate a full Gaussian process while drastically reducing computational complexity \citep{finley2019efficient}, which allows \texttt{spOccupancy} to fit computationally efficient models even for large spatial data sets. In this paper, we review SSOMs, MSOMs, and IOMs, describe the functionality of \texttt{spOccupancy}, and illustrate its features using two case studies on breeding birds in the USA.

\section{Models}\label{models}

\texttt{spOccupancy} fits nonspatial and spatial SSOMs, MSOMs, and IOMs (six different model structures; see Table \ref{tab:coreFunctions} for function names and descriptions). All models are fit in a Bayesian framework with priors and numerical algorithms implemented to maximize computational efficiency. See Supplemental Information S1.3 for a detailed discussion on the prior distributions and their default values in \texttt{spOccupancy}.  

\begin{table}[ht] % <--
  \begin{center}
    \caption{List of core functions in the \texttt{spOccupancy} package. The \texttt{PG} in model fitting function names refers to the \pg data augmentation approach \citep{polson2013} used to fit all occupancy models.}
    \label{tab:coreFunctions}
    \begin{tabular}{| l | l |}
      \hline
      Functionality & Description \\
      \arrayrulecolor{gray}\hline
      \textbf{Data simulation} & \\
      \arrayrulecolor{gray}\hline
      \texttt{simOcc} & Simulate single-species occupancy data \\
      \texttt{simMsOcc} & Simulate multi-species occupancy data \\
      \texttt{simIntOcc} & Simulate single-species occupancy data from multiple data sources \\
      \arrayrulecolor{gray}\hline
      \textbf{Model fitting} & \\
      \arrayrulecolor{gray}\hline
       \texttt{PGOcc} & Single-species occupancy model \\
      \texttt{spPGOcc} & Single-species spatial occupancy model \\
      \texttt{intPGOcc} & Single-species occupancy model with multiple data sources \\
      \texttt{spIntPGOcc} & Single-species spatial occupancy model with multiple data sources \\
      \texttt{msPGOcc} & Multi-species occupancy model \\
      \texttt{spMsPGOcc} & Multi-species spatial occupancy model \\
      \arrayrulecolor{gray}\hline
      \textbf{Model assessment} & \\
      \arrayrulecolor{gray}\hline
      \texttt{ppcOcc} & Posterior predictive check using Bayesian p-values \\
      \texttt{waicOcc} & Compute Widely Applicable Information Criterion \\
    \hline
    \end{tabular}
  \end{center}	 
\end{table}

\subsection{Single-species occupancy models (SSOMs)}

Let $z_{j}$ denote the true presence (1) or absence (0) of a species at site $j = 1, \dots, J$. The SSOM assumes $z_j$ arises from a Bernoulli process following

\begin{equation}\label{SSOM-Z}
  z_j \sim \text{Bernoulli}(\psi_j),
\end{equation}

where $\psi_j$ is the probability of occurrence at site $j$ \citep{mackenzie2002, tyre2003improving}. We model $\psi_j$ using a logit link following

\begin{equation}\label{SSOM-Psi}
  \text{logit}(\psi_j) = \bm{x}^\top_j\bm{\beta},
\end{equation}

where $\bm{\beta}$ is a vector of regression coefficients (including an intercept) that describe the effect of covariates $\bm{x}_j$ and the $^\top$ denotes transposition of column vector $\bm{x_j}$. 

To estimate $\psi_j$ while accounting for imperfect detection, $k = 1, \dots, K_j$ sampling replicates are obtained at each site $j$. We model the observed detection (1) or nondetection (0) of a study species during replicate visit $k$ at site $j$, denoted $y_{j, k}$, conditional on the true occupancy process, $z_j$, following 

\begin{equation}\label{SSOM-Y}
  y_{j, k} \sim \text{Bernoulli}(p_{j, k}z_j),
\end{equation}

where $p_{j, k}$ is the probability of detecting the species at site $j$ during visit $k$. Detection probability can vary by site and/or sampling covariates following

\begin{equation}\label{SSOM-P}
  \text{logit}(p_{j, k}) = \bm{v}^\top_{j, k}\bm{\alpha}, 
\end{equation}

where $\bm{\alpha}$ is a vector of regression coefficients (including an intercept) that describe the effect of site and/or observation covariates $\bm{v}_{j, k}$ on detection.

We complete the Bayesian specification of the model by assigning Gaussian priors to the occurrence ($\bm{\beta}$) and detection ($\bm{\alpha}$) regression coefficients (including the intercepts).  See Supplemental Information S2.2 for further model details. The \texttt{spOccupancy} function \texttt{PGOcc} fits SSOMs.

\subsection{Spatial single-species occupancy models (SSOMs)}

We extend the previous SSOM to account for residual spatial variation in species occurrence. Let $\bm{s}_j$ denote the geographical coordinates of site $j$ for $j = 1, \dots, J$. In the spatial SSOM, we include $\bm{s}_j$ directly in the notation of spatially-indexed variables to indicate the model is spatially-explicit. More specifically, the occurrence probability at site $j$ with coordinates $\bm{s}_j$, $\psi(\bm{s}_j)$, now takes the form 

\begin{equation}\label{spatial-SSOM-psi}
\text{logit}(\psi(\bm{s}_j)) = \bm{x}(\bm{s}_j)^\top\bm{\beta} + \text{w}(\bm{s}_j), 
\end{equation}

where $\text{w}(\bm{s}_j)$ is a realization from a zero-mean spatial Gaussian process. In particular, we assume that

\begin{equation}\label{singleSpeciesW}
	\text{\textbf{$\text{w}(\bm{s})$}} \sim N(\bm{0}, \Sigma(\bm{s}, \bm{s}', \bm{\theta})),
\end{equation}

where $\Sigma(\bm{s}, \bm{s}', \bm{\theta})$ is a $J \times J$ covariance matrix that is a function of the distances between any pair of site coordinates $\bm{s}$ and $\bm{s}'$ and a set of parameters ($\bm{\theta}$) that govern the spatial process according to a spatial correlation function. \texttt{spOccupancy} supports four spatial correlation functions: exponential, spherical, Gaussian, and Mat\'ern (see Chapters 1 and 3 in \cite{banerjee2003} for correlation function details). For the exponential, spherical, and Gaussian functions, $\bm{\theta} = \{\sigma^2, \phi\}$, where $\sigma^2$ is the spatial variance parameter and $\phi$ is a spatial decay parameter, while the Mat\'ern specification additionally includes a spatial smoothness parameter, $\nu$. As a result of the additional smoothness parameter, the Mat\'ern correlation function is the most flexible out of the four, but using this function may require comparatively more data. Generally, these functions can all adequately accommodate spatial autocorrelation, and we recommend using the Widely Applicable Information Criterion (WAIC; \citealt{watanabe2010}) and k-fold cross-validation to select among different correlation functions (see `Implementation and Usage of \texttt{spOccupancy}` section). We assign an inverse-Gamma prior to the spatial variance parameter ($\sigma^2$) and uniform priors to the spatial decay ($\phi$) and smoothness ($\nu$) parameters. The remainder of the model follows the nonspatial SSOM (Supplemental Information S2.3). The \texttt{spOccupancy} function \texttt{spPGOcc} fits spatial SSOMs.  

\subsection{Multi-species occupancy models (MSOMs)}

As an extension to the SSOM, consider a data set $y_{i, j, k}$ denoting the detection or nondetection of species $i$ at site $j$ during replicate $k$ for $i = 1, \dots, N$ species. The ecological process of interest is $z_{i, j}$, the true presence or absence status of species $i$ at site $j$. Following \cite{dorazio2005}, $z_{i, j}$ and $y_{i, j, k}$ are modeled according to Equations \ref{SSOM-Z} and \ref{SSOM-Y} with all detection and occupancy parameters now varying by species. Species-specific regression coefficients for occupancy ($\bm{\beta}_i$) and detection ($\bm{\alpha}_i$) are treated as random effects arising from community-level normal distributions, which leads to greater precision of species-specific effects (particularly for rare species) and facilitates estimation of biodiversity metrics \citep{zipkin2009impacts}. For example, the species-specific occurrence intercept, $\beta0_{i}$, is modeled according to  

\begin{equation}\label{communityRegression}
\beta0_{i} \sim \text{Normal}(\mu_{\beta0}, \tau^2_{\beta0}),
\end{equation}

where $\mu_{\beta0}$ is the community-level occurrence intercept and $\tau^2_{\beta0}$ is the variance of the intercept among species in the community. Models for all parameters in $\bm{\beta}_i$ and $\bm{\alpha}_i$ are defined analogously. We assign normal priors to the community level occurrence ($\bm{\mu_{\beta}}$) and detection ($\bm{\mu_{\alpha}}$) regression coefficients, and inverse-Gamma priors to each of the occurrence ($\bm{\tau^2}_{\bm{\beta}}$) and detection ($\bm{\tau^2}_{\bm{\alpha}}$) variance parameters (Supplemental Information S2.4). The \texttt{spOccupancy} function \texttt{msPGOcc} fits MSOMs. 

\subsection{Spatial multi-species occupancy models (MSOMs)}

The spatial MSOM is identical to the MSOM except the occurrence probability for each species also includes a species-specific spatial Gaussian process ($\text{\textbf{w}}_i(\bm{s})$), along with associated parameters ($\bm{\theta}_i$), analogous to the spatial SSOM (Equation \ref{spatial-SSOM-psi}). The spatial parameters are estimated individually for each species using inverse-Gamma priors for the spatial variance parameters and uniform priors for the spatial range and smoothness parameters (Supplemental Information S2.5). The function \texttt{spMsPGOcc} fits spatial MSOMs. 

\subsection{Integrated occupancy models (IOMs)}

Model-based integration of multiple data types has become common as the number of available data sources has increased \citep{miller2019recent}. When integrating multiple detection-nondetection data sources in an IOM, each data source has its own unique detection model defined by Equations \ref{SSOM-Y} and \ref{SSOM-P} that are conditional on a shared latent occupancy process defined by Equations \ref{SSOM-Z} and \ref{SSOM-Psi}. This joint-likelihood approach enables explicit estimation of different covariate effects (and intercepts) on the detection processes of each data source (Supplemental Information S2.6). The \texttt{spOccupancy} function \texttt{intPGOcc} fits IOMs. 

\subsection{Spatial integrated occupancy models (IOMs)}

The spatial IOM is identical to the IOM except the latent occurrence probability includes a spatial random intercept following Equation \ref{spatial-SSOM-psi} (Supplemental Information S2.7). The function \texttt{spIntPGOcc} fits spatial IOMs. 

\section{Computational Advances in \texttt{spOccupancy}}

Bayesian occupancy models using a logit link function are often slow and inefficient in standard Bayesian software packages \citep{clark2019}, which arises from a need to use inefficient algorithms to estimate the occurrence and detection regression parameters. We avoid this computational burden by using \pg data augmentation \citep{polson2013}, a statistical approach in which we introduce two sets of latent auxiliary variables that follow a \pg distribution, which results in an efficient Gibbs update for the occurrence and detection regression parameters in occupancy models with a logit link that is orders of magnitude faster than traditional algorithms \citep{clark2019}. See Supplemental Information S1.1 for details. 

Spatial models are a notorious computational bottleneck once the number of sites ($J$) becomes even moderately large. We provide users the option to fit all spatial occupancy models in \texttt{spOccupancy} using a Nearest Neighbor Gaussian Process (NNGP; \citealt{datta2016hierarchical}), which enables spatially-explicit occupancy modeling of data sources comprising locations in the tens to hundreds of thousands. The NNGP uses local information from a reduced set of nearest neighbors to provide inferences that are nearly indistinguishable from the full Gaussian process. Fifteen neighbors is often sufficient, although as few as five neighbors may be adequate for certain data sets with long-range spatial dependence \citep{datta2016hierarchical}. See Supplemental Information S1.2 for additional details.

We wrote all occupancy models fit by \texttt{spOccupancy} in \texttt{C/C++} using \texttt{R}'s foreign language interface. Our code draws heavily from the computational technology employed in the \texttt{spNNGP} package \citep{finley2020spnngp}. See Supplemental Information S2 for complete details on the Gibbs samplers used in \texttt{spOccupancy}.

\section{Implementation and Usage of \texttt{spOccupancy}}

Here we briefly describe functionality for the five main tasks performed by \texttt{spOccupancy} (Table~\ref{tab:coreFunctions}). See the package vignette (Supplemental Information S3), the package website (\url{https://www.jeffdoser.com/files/spoccupancy-web/}), and the \texttt{R} package documentation for additional details and examples.

\textit{1. Data simulation}. The functions \texttt{simOcc}, \texttt{simMsOcc}, and \texttt{simIntOcc} simulate data under the SSOM, MSOM, and IOM frameworks. All simulation functions include arguments to optionally simulate data with spatial random effects in the occurrence portion of the model.  

\textit{2. Model fitting}. Each of the model fitting functions was described previously (Section \ref{models}). Functions for SSOMs and MSOMs allow for the inclusion of random intercepts in the occurrence and detection portion of the occupancy model (e.g., random effects for some habitat classification, random observer effects). Users can specify each parameter's prior distribution to yield vague or informative priors as desired, with the default being weakly informative priors (See Supplemental Information S1.3).

\textit{3. Model validation and comparison}. The \texttt{spOccupancy} function \texttt{ppcOcc} performs posterior predictive checks on all \texttt{spOccupancy} model objects, with options to calculate Bayesian p-values as a simple assessment of model fit. For model selection and assessment, the function \texttt{waicOcc} computes the WAIC \citep{watanabe2010}. Alternatively, users can perform k-fold cross-validation using the \texttt{k.fold} argument in all \texttt{spOccupancy} model fitting functions. We use the model deviance as a scoring rule for k-fold cross-validation \citep{hooten2015guide}. 

\textit{4. Posterior summaries}. All posterior samples are returned as \texttt{coda::mcmc} objects \citep{coda}. We include \texttt{summary} functions for all \texttt{spOccupancy} model objects, which print concise summaries (e.g., posterior quantiles, posterior means) of the posterior distributions for estimated parameters as well as the Gelman-Rubin diagnostic (Rhat; \citealt{brooks1998}) and effective sample size for convergence diagnostics.  

\textit{5. Prediction}. We implement a \texttt{predict} function for all \texttt{spOccupancy} model objects to yield predictions of latent occurrences and occurrence probabilities across a set of locations (given covariate values and spatial coordinates), which may or may not include a subset (or all) of the surveyed locations. The resulting object consists of posterior predictive distributions which can be used to provide maps of occurrence probability (i.e, species distribution maps) and fully propagated estimation uncertainty. We additionally allow users to predict detection probability across a user-specified range of covariate values.

\section{Case Studies}

We demonstrate \texttt{spOccupancy} functionality  with two case studies on forest breeding birds in the eastern USA. See Supplemental Information S1 for full case study details and an analysis of three simulated data sets across $J = 40,000$ locations using a spatial IOM. 

\subsection{Black-throated Green Warbler in eastern USA (SSOM)}

For our first example, we estimated occurrence of Black-throated Green Warbler (\textit{Setophaga virens}) across the eastern USA in 2018. We used data from the North American Breeding Bird Survey (BBS; \citealt{pardieck2020north}), where observers perform roadside surveys at 50 stops along $\sim$3000 routes each year distributed across the USA and Canada. We modeled route-level occurrence as a function of local forest cover (linear) and elevation (linear and quadratic) and modeled detection as a function of day of survey (linear and quadratic), time of day (linear), and a random observer effect. All variables were standardized to have mean 0 and standard deviation 1. We fit a non-spatial and a spatial SSOM using \texttt{PGOcc} and \texttt{spPGOcc}, respectively. We fit the spatial SSOM using a full Gaussian process and also with a NNGP with 15 neighbors to assess the computational benefits provided by the NNGP. For both spatial models we used an exponential correlation function. 

\begin{figure}[ht]
    \centering
    \includegraphics[width=15cm]{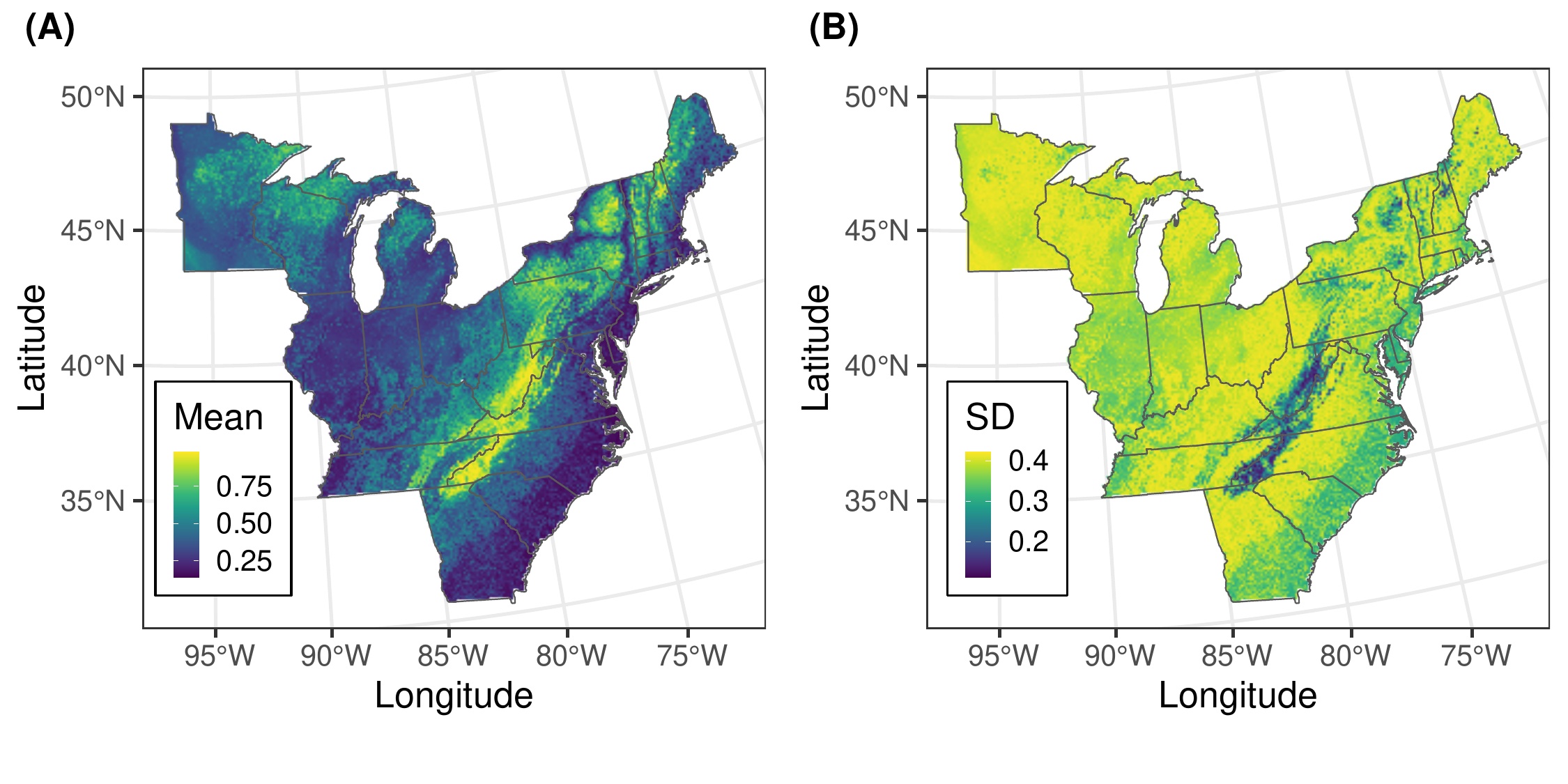}
    \caption{Predicted occurrence probability mean (A) and standard deviation (B) for Black-throated Green Warbler across the eastern US in 2018 as estimated using a spatial SSOM.}
    \label{fig:btbwFigure}    
\end{figure}

The spatial SSOMs outperformed the nonspatial SSOM according to both the WAIC and 10-fold cross-validation (Table \ref{tab:bbsTable}). The NNGP spatial model provided massive increases in computing efficiency compared to the full Gaussian process spatial model (Table~\ref{tab:bbsTable}) with only small differences in parameter estimates and effective sample sizes. Predicted occurrence probabilities from the NNGP spatial SSOM indicated high occurrence of Black-throated Green Warbler across the Appalachian mountains, the northeastern states, and a portion of the northern Midwest (Figure \ref{fig:hbefFigure}).  

\begin{table}[ht] % <--
  \begin{center}
  \caption{Posterior median estimates (95\% credible intervals), WAIC, 10-fold cross-validation (CV) deviance, and run time for a SSOM and spatial SSOMs using a NNGP and a full Gaussian process (GP) for the Black-throated Green Warbler case study. Run times are times to complete 50,000 MCMC samples.} 
  \label{tab:bbsTable}
  \begin{tabular}{|d | a | d | a|}
    \hline
    & Nonspatial & Spatial (NNGP) & Spatial (GP) \\
    \hline
    \textbf{Occurrence} & & & \\
    Intercept & -1.43 (-1.74, -1.13) & -1.28 (-3.98, 0.96) & -1.61 (-3.94, 0.94)\\
    Linear Elevation & 1.38 (1.06, 1.73) & 2.13 (1.22, 3.27) & 1.95 (1.05, 3.00)\\
    Quadratic Elevation & 0.26 (-0.08, 0.61) & 0.16 (-0.32, 0.82) & 0.21 (-0.30, 0.90) \\
    Forest Cover & 1.15 (0.89, 1.44) & 1.11 (0.58, 1.84) & 1.06 (0.52, 1.68) \\
    Spatial Variance & - & 23.57 (9.77, 51.89) & 21.44 (8.53, 45.46) \\
    Spatial Range & - & 0.0021 (0.0013, 0.0039) & 0.0022 (0.0013, 0.0041) \\
    \hline
    \textbf{Detection} & & & \\
    Intercept & -0.46 (-0.90, -0.08) & -0.14 (-0.44, 0.15) & -0.14 (-0.45, 0.16) \\
    Linear Day & -0.15 (-0.40, 0.10) & -0.14 (-0.37, 0.08) & -0.14 (-0.34, 0.08) \\
    Quadratic Day & 0.062 (-0.17, 0.30) & 0.033 (-0.17, 0.24) & 0.040 (-0.16, 0.24) \\
    Time of Day & -0.01 (-0.32, 0.29) & -0.010 (-0.25, 0.24) & -0.019 (-0.27, 0.24) \\
    Observer Variance & 2.49 (1.55, 3.70) & 1.60 (1.01, 2.34) & 1.58 (0.99, 2.37) \\
    \hline
    WAIC & 2401.16 & 2117.01 & 2118.95 \\
    CV Deviance & 3317.39 & 2504.03 & 2537.67 \\
    Run Time (min) & 3.17 & 7.16 & 1670.73 \\
    \hline
  \end{tabular}
  \end{center}
\end{table}

\subsection{Foliage-gleaning birds in Hubbard Brook (MSOM)}

For our second case study, we estimated species richness for a community of twelve foliage-gleaning birds in 2015 in the Hubbard Brook Experimental Forest in New Hampshire, USA. Data were collected using standard point count surveys at 373 sites three times during the breeding season. We included time of day (linear) and day of year (linear and quadratic) as fixed effects in the detection portion of the model, and specified linear and quadratic effects of elevation as occurrence predictors. All variables were standardized to have mean 0 and standard deviation 1. We fit a series of nonspatial and spatial MSOMs using \texttt{msPGOcc} and \texttt{spMsPGOcc} where we compared the benefits of including elevation as an occurrence predictor and spatial random effects using an exponential correlation function. We fit all spatial models using an NNGP with five neighbors.   

\begin{table}[ht] % <--
  \begin{center}
  \caption{Candidate model community-level posterior median estimates (95\% credible intervals), WAIC, and four-fold cross-validation (CV) deviance in the foliage-gleaning bird case study in the Hubbard Brook Experimental Forest. $\mu_{\beta_0}$ is the community-level intercept, $\mu_{\beta_1}$ and $\mu_{\beta_2}$ are the linear and quadratic effects of elevation, respectively, and $\tau^2_{\beta, 0}$, $\tau^2_{\beta, 1}$, and $\tau^2_{\beta, 2}$ are community-level variances for the intercept, linear, and quadratic elevation effects across species, respectively. Run times are minutes to complete 150,000 MCMC iterations.}
  \label{tab:hbefTable}
  \begin{tabular}{|d | a | d | a | d | }
    \hline
    & Intercept & Intercept + & Elevation & Elevation +\\
    & & Spatial & & Spatial \\
    \hline
    $\mu_{\beta_0}$ & 0.37 (-0.82, 1.55) & 0.38 (-1.44, 2.19) & 0.43 (-1.28, 2.23) & 0.42 (-1.59, 2.36)\\
    $\mu_{\beta_1}$ & - & - & 0.26 (-0.79, 1.24) & 0.29 (-1.08, 1.60)\\
    $\mu_{\beta_2}$ & - & - & -0.21 (-0.68, 0.30) & -0.26 (-0.90, 0.43)\\
    $\tau^2_{\beta, 0}$ & 4.51 (1.39, 13.10) & 12.99 (3.61, 39.51) & 11.67 (4.21, 30.83) & 18.10 (6.60, 45.84)\\
    $\tau^2_{\beta, 1}$ & - & - & 3.21 (1.02, 8.56) & 5.70 (1.66, 15.18) \\
    $\tau^2_{\beta, 2}$ & - & - & 0.60 (0.15, 1.76) & 1.00 (0.22, 3.14) \\
    \hline
    WAIC & 9714.31 & 9057.95 & 9195.61 & 9043.74 \\
    CV Deviance & 10137.57 & 10091.36 & 9989.48 & 10063.4 \\
    Run Time & 20.56 & 42.69 & 20.90 & 42.74 \\
    \hline
  \end{tabular}
  \end{center}
\end{table}

\begin{figure}
    \centering
    \includegraphics[width=12cm]{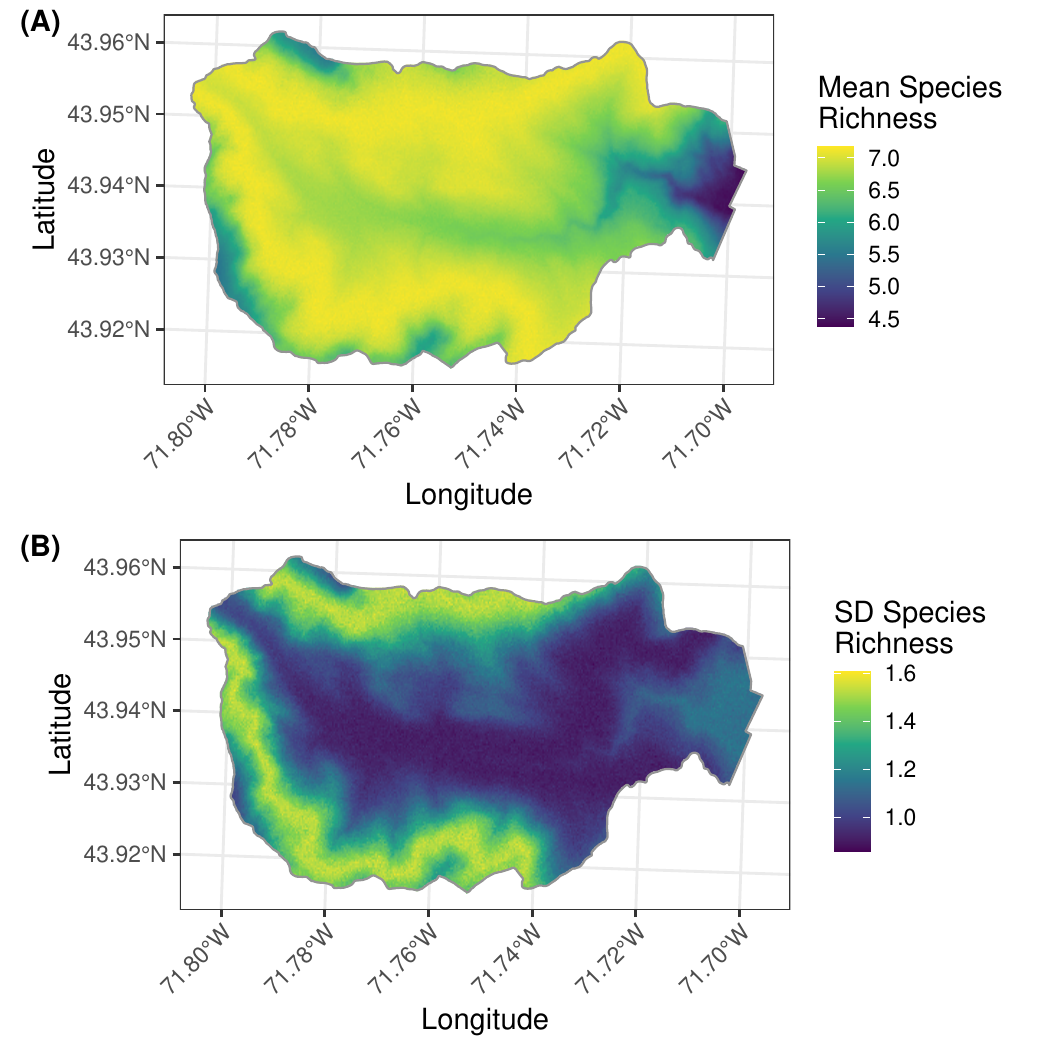}
    \caption{Estimate of species richness and the associated uncertainty of the community of twelve foliage-gleaning birds across the Hubbard Brook Experimental Forest from a non-spatial MSOM. Panel (A) shows posterior means and Panel (B) shows posterior standard deviations.}
    \label{fig:hbefFigure}
\end{figure}

The spatial MSOM that included linear and quadratic elevation covariates on occurrence was the best performing model according to the WAIC, but the nonspatial MSOM with elevation performed best according to four-fold cross-validation (Table \ref{tab:hbefTable}). The spatially-explicit intercept-only model outperformed the nonspatial intercept-only model according to both criteria, which altogether indicates there is spatial variation in foliage-gleaning bird occurrence across the study region, but that this variation is fairly well-explained by the important elevational gradient in the study area. We predicted species richness across Hubbard Brook using the nonspatial MSOM with the elevation covariate, which revealed moderate variation across Hubbard Brook with areas of low and high elevation having on average lower richness than moderate elevations (Figure \ref{fig:hbefFigure}A) and high uncertainty along the edges of the forest, which closely corresponds to high elevation areas (Figure \ref{fig:hbefFigure}B, Supplemental Information S1.S4 Figure S1).

\section{Conclusions and Future Directions}

Our \texttt{spOccupancy} \texttt{R} package fits spatially-explicit single-species, multi-species, and integrated occupancy models for potentially massive data sets. The package includes functions for data simulation, model fitting, model validation, model comparison, and out-of-sample prediction. The package vignette (Supplemental Information S3) and website (\url{https://www.jeffdoser.com/files/spoccupancy-web/}) contain full details and examples on all \texttt{spOccupancy} model functions. We are currently working on including the following extensions within the package: (1) dynamic occupancy models \citep{mackenzie2003estimating}; (2) spatially varying coefficients (SVCs; \citealt{finley2011comparing}) in the occurrence model; (3) multi-species integrated occupancy models \citep{doser2022integrated}. We expect \texttt{spOccupancy} will serve as a user-friendly tool for ecologists and conservation practitioners to account for detection biases and spatial autocorrelation using large data sets (e.g., hundreds of thousands of locations) in assessments of species distributions and community patterns across broad spatial regions, an increasingly important objective in species distribution modeling applications. 

\section{Data Availability Statement}

The package \texttt{spOccupancy} is available on the Comprehensive R Archive Network (CRAN; \url{https://cran.r-project.org/web/packages/spOccupancy/index.html}). Data and code used in the examples are available on GitHub (\url{https://github.com/doserjef/Doser_etal_2021_spOccupancy}) and will be posted on Zenodo upon acceptance. 

\section*{Authors' Contributions}

JWD developed the package with guidance from AOF. JWD performed analyses and led writing of the manuscript with critical insights from MK, EFZ, and AOF. All authors gave final approval for publication. 
 
\section*{Acknowledgements}

We declare no conflicts of interest. We thank Sam Ayebare, Courtney Davis, and Gabriela Quinlan for insightful comments on the package, as well as Scott Sillet and Mike Hallworth for providing the Hubbard Brook data. This work was supported by National Science Foundation grants DMS-1916395, EF-1253225, and DBI-1954406.

\bibliographystyle{apalike}

\end{document}